\documentclass[sigconf]{acmart}
\usepackage{tabularx}
\newcolumntype{Y}{>{\centering\arraybackslash}X}

\usepackage{algorithm}
\usepackage{algpseudocode}
\usepackage{multirow}
\usepackage{subcaption}
\usepackage{cleveref}
\algdef{S}[FOR]{ForEach}[1]{\algorithmicforeach\ #1\ \algorithmicdo}

\copyrightyear{2024} 
\acmYear{2024} 
\setcopyright{acmcopyright}
\acmConference[CIKM '24]{Proceedings of the 31st ACM International Conference on Information and Knowledge Management}{May 7--11, 2024}{Atlanta, GA, USA}
\acmBooktitle{Proceedings of the 31st ACM International Conference on Information and Knowledge Management (CIKM '24), May 7--11, 2024, Atlanta, GA, USA}
\acmPrice{15.00}
\acmDOI{10.1145/3511808.3557065}
\acmISBN{978-1-4503-9236-5/24/10}
\begin{document}

\title{UnifiedRL: A Reinforcement Learning Algorithm Tailored for Multi-Task Fusion in Large-Scale Recommender Systems}

\author{Peng Liu}
\email{liupengvswww@gmail.com}
\orcid{0009-0000-7271-4721}
\affiliation{%
  \institution{Tencent Inc.}
  \city{Beijing}
  \country{China}
}

\author{Cong Xu}
\email{congcxu@tencent.com}
\affiliation{%
  \institution{Tencent Inc.}
  \city{Beijing}
  \country{China}
}

\author{Ming Zhao}
\email{marcozhao@tencent.com}
\affiliation{%
  \institution{Tencent Inc.}
  \city{Beijing}
  \country{China}
}

\author{Jiawei Zhu}
\email{erickjwzhu@tencent.com}
\affiliation{%
  \institution{Tencent Inc.}
  \city{Beijing}
  \country{China}
}

\author{Bin Wang}
\email{hillmwang@tencent.com}
\affiliation{%
  \institution{Tencent Inc.}
  \city{Beijing}
  \country{China}
}

\author{Yi Ren}
\email{henrybjren@tencent.com}
\affiliation{%
  \institution{Tencent Inc.}
  \city{Beijing}
  \country{China}
}

\renewcommand{\shortauthors}{Peng Liu et al.}

\begin{abstract}
As the last pivotal stage of Recommender System (RS), Multi-Task Fusion (MTF) is responsible for combining multiple scores outputted by 
Multi-Task Learning (MTL) model into a final score to maximize user satisfaction. 
Recently, to optimize long-term user satisfaction, Reinforcement Learning (RL) is used for MTF in RSs. 
However, the existing offline RL algorithms used for MTF have the following severe problems: 
a) To avoid Out-of-Distribution (OOD), their constraints are overly strict, which seriously damage performance; 
b) They are unaware of the exploration policy used to collect training data, only suboptimal policy can be learned; 
c) Their exploration policies are inefficient and hurt user experience. 

To solve the above problems, we propose an innovative method called UnifiedRL tailored for MTF in large-scale RSs. 
UnifiedRL seamlessly integrates offline RL model with its custom exploration policy to relax overly strict constraints, 
which is different from existing RL-MTF methods and significantly improves performance. 
In addition, compared to existing exploration policies, UnifiedRL's custom exploration policy is highly efficient, 
enabling frequent online exploration and offline training iterations, which further improves performance. 
Extensive offline and online experiments are conducted in a large-scale RS. 
The results demonstrate that UnifiedRL outperforms other existing MTF methods remarkably, 
achieving a $+4.64\%$ increase in user valid consumption and a $+1.74\%$ increase in user duration time. 
To the best of our knowledge, UnifiedRL is the first RL algorithm tailored for MTF in RSs and 
has been successfully deployed in multiple large-scale RSs since June 2023, yielding significant benefits.
\end{abstract}


\maketitle

\section{Introduction}

\label{sec:intro}
Recommender System (RS) \cite{ref1, ref2}, which provides personalized recommendation service based on user preference are 
widely used in various platforms, including short video platforms \cite{ref3, ref7, ref14}, video platforms \cite{ref4, ref5}, 
e-commerce platforms \cite{ref6, ref8, ref9, ref10, ref11} and social networks \cite{ref12, ref13}. 
Industrial RS mainly includes three stages: candidate generation, ranking and Multi-Task Fusion (MTF) \cite{ref4, ref15}. 
During candidate generation, thousands of candidates are selected from millions or even billions of items. 
Ranking typically uses a Multi-Task Learning (MTL) \cite{ref4, ref8, ref16, ref17, ref18, ref40} model to estimate the scores of 
various user behaviors such as valid click, watching time, fast slide, like and sharing. 
Finally, a MTF model combines multiple scores outputted by MTL model into a final score to produce the final ranking of 
candidates \cite{ref15}, which decides the ultimate recommendation results. 
However, there is little valuable research on RS MTF until now.

\begin{figure}[hbtp!]
  \centering
  \includegraphics[width=0.95\linewidth]{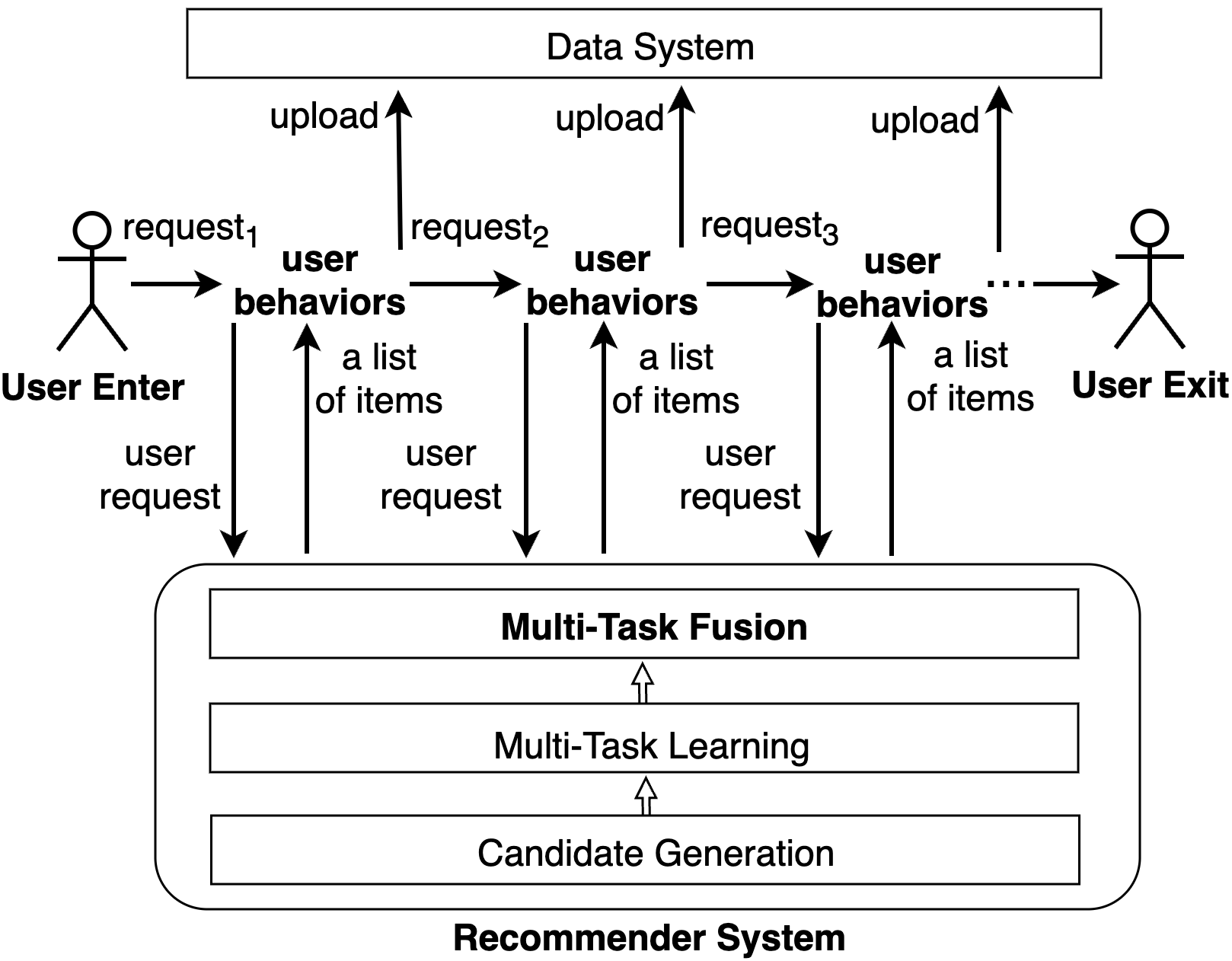}
  \caption{The interactions between the user and the RS within a recommendation session.}
  \label{fig:session}
\end{figure}

The goal of MTF in RSs is to maximize user satisfaction, which is commonly evaluated by the weighted sum of the user's various feedbacks 
including watching time, valid click, like, sharing and other behaviors within a single recommendation or a recommendation session. 
A recommendation session is defined as the process from when a user starts accessing RS to leaving, 
which includes one or more consecutive interactions, as shown in Figure \ref{fig:session}. 
Early works such as Grid Search \cite{ref36} and Bayesian Optimization \cite{ref19} ignore user preference and merely 
generate the same fusion weights for all users. 
Evolution Strategy (ES) \cite{ref20, ref21, ref22} can produce personalized fusion weights but is limited by its learning pattern \cite{ref23}. 
Furthermore, all the methods mentioned above only focus on the reward of current recommendation but ignore long-term rewards.

In personalized RSs such as YouTube, TikTok, Shorts and Kwai, the current recommendation has a significant impact on its subsequent recommendations, 
especially within a recommendation session. Therefore, both the reward of current recommendation and 
the rewards of subsequent recommendations within a session need to be considered together. 
Recently, a few studies \cite{ref15} have used offline Reinforcement Learning (RL) algorithms \cite{ref26, ref27, ref28, ref29} to 
search for the optimal fusion weights to maximize long-term rewards. 
However, they suffer from the following serious problems \cite{ref15, ref26, ref27, ref28, ref29, ref30, ref31}: 
a) To avoid Out-of-Distribution (OOD) problem, their constraints are excessively strict, which significantly impairs their performance.
b) Online exploration and offline model training are two independent processes, offline RL algorithms are unaware of the exploration policy 
 behind training data, therefore only suboptimal policy can be learned.
c) The existing exploration policies are inefficient and negatively impact user experience.

To solve the aforementioned problems, we propose a novel method UnifiedRL, specifically designed for MTF based on the characteristics of RSs.
First, UnifiedRL integrates offline RL model with its custom exploration policy. 
During offline training, the distribution of training data generated by its exploration policy can be directly obtained. 
Therefore, the overly strict constraints used to avoid OOD problem can be relaxed, which significantly improves UnifiedRL's performance. 
Second, we design a simple yet extremely efficient exploration policy, which not only accelerates model iteration but also 
alleviates negative impacts on user experience. 
Finally, we adopt a progressive training mode to further improve performance with the help of its tailored exploration policy, 
which enables the target policy to swiftly converge toward the true optimal policy through 
frequent iterations of online exploration and offline model training.

In this paper, our contribution can be summarized as follows:
\begin{itemize}
    \item We highlight that existing RL methods used for MTF overlook the unique characteristics of RSs, which suffer from several severe problems,
    resulting in the learning of only suboptimal policies and inefficiency exploration.

    \item We propose UnifiedRL tailored for MTF in large-scale RSs, which combines the offline RL algorithm and the custom efficient exploration policy 
    into a unified framework, significantly improving model performance. 
    Moreover, with the help of UnifiedRL's custom exploration policy, a progressive training mode is adopted to further enhance performance.

    \item Extensive offline and online experiments demonstrate that UnifiedRL significantly outperforms other existing MTF methods,
    achieving a $+4.64\%$ increase in user valid consumption and a $+1.74\%$ increase in user duration time. 

    \item To the best of our knowledge, UnifiedRL is the first RL algorithm specifically designed for MTF in RSs, 
    and it has already been successfully applied in several large-scale RSs. 
    In addition, UnifiedRL has also been used in search engines and advertising systems.
\end{itemize}


\section{Related Work}
\label{related_work}
In RS, Multi-Task Fusion (MTF) \cite{ref4, ref15} is responsible for merging multiple predictions into a single score, 
thereby determining the final recommendation results, which is critical for RS. 
However, the valuable research on MTF in RS to optimize user satisfaction is not much. 
Initially, Grid Search \cite{ref36} is used to find the optimal fusion weights by searching through a candidate set 
composed of numerous parameter combinations. 
Later, Bayesian Optimization \cite{ref19, ref37} is proposed to accelerate the parameter search process. 
The primary drawback of these two methods is that they ignore user preference and produce the same fusion weights for all users. 
In addition, these methods are inefficient and are rarely used in industrial RSs today.
ES \cite{ref20, ref21, ref22} takes user preference features as model input to output personalized fusion weights for different users. 
ES updates its model parameters through a process of alternating mutation and selection. 
Due to this learning pattern, the number of ES model parameters must be small, which limits its performance \cite{ref23}. 
Furthermore, all the methods mentioned above focus only on instant reward and ignore long-term reward.

In personalized RSs such as YouTube, TikTok, Kwai, and Little Red Book \cite{ref3, ref7, ref14}, 
current recommendation has an obvious influence on subsequent recommendations. 
Therefore, in recent years, some works try to search for optimal fusion weights via RL to maximize long-term reward. 
Compared with the aforementioned methods, RL considers the cumulative reward within a session and 
recommends items that not only satisfy users in current recommendation but also promote positive long-term engagement. 
It should be emphasized that although RL has also been applied to solve other problems in RSs, 
this paper focuses on RL algorithms for MTF to optimize user satisfaction and does not discuss other topics, 
such as learnable fusion formula \cite{ref44} or user retention optimization based on RL \cite{ref45}.

\cite{ref38} uses an offline RL algorithm to search for the optimal weights between the predicted click-through rate and 
the bid price of an advertiser. 
\cite{ref15} proposes BatchRL-MTF to find out the optimal fusion weights, which has been successfully deployed in multiple RSs at Tencent. 
Some teams replace the BCQ algorithm \cite{ref28} in BatchRL-MTF with other RL algorithms, such as DDPG \cite{ref27}, CQL+SAC \cite{ref30, ref39}, 
IQL \cite{ref31}, etc. 
However, the existing works on RL-MTF do not take into account the unique characteristics of RS and have the following serious problems: 
First, to avoid OOD problem, their constraints are too strict, 
significantly hurting their performance. Take BCQ as an example, the action generated by target policy for a given state must closely 
resembles the distribution of action for this state contained in the exploration data. 
Second, online exploration and offline model training are two independent processes. 
The existing offline RL algorithms are unaware of the exploration policy used for 
generating training data, so the optimal policy cannot be learned. 
Third, the exploration policies of existing methods ignore the study of efficiency in RSs, 
resulting in low efficiency and negative impact on user experience.


\section{Problem Definition}
\label{sec:problem}
This section gives the problem definition of RL-MTF in RS.
As mentioned above, the current recommendation has an evident influence on subsequent recommendations within a recommendation session. 
At each time step $t$, after RS receives a user's request, the following steps are performed:
First, thousands of candidates are selected from millions of or even more items.
Second, the MTL model predicts scores for multiple user behaviors for each candidate.
Third, the MTF model generates fusion weights to combine multiple scores output by the MTL model into a final score using Eq. 1. 
It is a commonly used fusion formula in RSs, where $pred\_score_{i}$ corresponds to the predicted values of MTL model, 
$power_i$ and $bias_i$ correspond to different elements of the action vector outputted by the RL-MTF model.
Finally, a list of items is generated and sent to the user, and the user's feedbacks is reported to the platform's data system.

\begin{equation}
    final\_score = \prod\limits_{i=1}^{k}\left(pred\_score_{i} + bias_{i}\right)^{power_{i}} \,.
\end{equation}

We model the above fusion problem within a recommendation session as a Markov Decision Process (MDP). 
In this MDP, RS acts as an agent that interacts with a user (environment) and makes sequential recommendations, 
aiming to maximize the cumulative reward within a session. 
The MDP framework has the following key components \cite{ref15, ref26, ref27, ref28, ref29}:

• \textbf{State Space ($\mathcal{S}$):} \textbf{$\mathcal{S}$} is a set of state $s$. The features of a state $s$ mainly include 
user profile feature (e.g., age, gender, top $k$ interests, flush num, etc.), 
user history behavior sequence (e.g., valid click, like, sharing, etc.) and statistical features.

• \textbf{Action Space ($\mathcal{A}$):} \textbf{$\mathcal{A}$} is a set of action $a$ generated by RL model. 
In the context of RS MTF problem, action $a$ is a fusion weight vector 
($a_{1}$, . . ., $a_{2k}$), of which each element corresponds to different power or bias term in Eq. 1.

• \textbf{Reward ($\mathcal{R}$):} After RS takes an action $a_{t}$ at state $s_{t}$ and sends a list of items to a user, 
the user's various behaviors to those items will be reported and 
the instant reward $r(s_{t}, a_{t})$ will be calculated based on these behaviors, as shown in Eq. 2.

• \textbf{Transition Probability ($\mathcal{P}$):} $p(s_{t+1}|s_{t}, a_{t})$ represents the likelihood of transitioning from 
state $s_{t}$ to state $s_{t+1}$ when action $a_{t}$ is taken. 
As mentioned above, a state $s_t$ mainly includes user profile feature and user history behavior sequence, 
so the next state $s_{t+1}$ depends on user feedbacks at time step $t$ and is deterministic.

• \textbf{Discount Factor ($\gamma$):} \textbf{$\gamma$} determines how much weight the agent assigns to future rewards 
compared with instant reward, $\gamma$ $\in$ $[0, 1]$. It is used to calculate the cumulative reward within a session, 
using the instant reward and subsequent rewards, as shown in Eq. 3.


\section{Method}
\label{sec:method}

\subsection{Reward Function}
\label{sec:reward_fun}
To evaluate instant satisfaction, we define instant reward function, as shown in Eq. 2, 
where $w_{i}$ is the weight of behavior $\upsilon_{i}$ and $k$ represents the number of user behavior types. 
It should be noted that the instant reward is calculated for the entire recommendation list, which contains $l$ items.
In our recommendation scenario, user behaviors ($\upsilon_{1}$, ..., $\upsilon_{k}$) 
contains watching time, valid consumption (watching a video longer than 10 seconds) and 
interaction behaviors such as like, sharing, collecting, etc. 
By analyzing the correlations between different user behaviors and long-term satisfaction, 
we set different weights for these behaviors. 

\begin{equation}
    r \left(s_{t}, a_{t}\right) = \sum\limits_{j=1}^{l}\sum\limits_{i = 1}^{k}w_{i} * \upsilon _{i,j} \,. 
\end{equation}

The calculation of RL cumulative reward, denoted as $G_t$, is presented in Eq. 3. 
$G_t$ represents the cumulative reward starting from time step $t$. 
$r(s_{t+i}, a_{t+i})$ denotes the instant reward received at step $t + i$. 
$\gamma$ is the discount factor and $T$ indicates the terminal time step.

\begin{equation}
G_t = \sum_{i=0}^{T-t} \gamma^i * r(s_{t+i}, a_{t+i}) \,. 
\end{equation}

\subsection{Online Exploration}
\label{sec:online_exploration}
Before training, a large amount of exploration data must first be collected, 
which has a critical impact on model performance. 
However, existing works on RL-MTF \cite{ref15, ref26, ref32} ignore exploration efficiency and potential negative impact on users,
which is very important for large-scale RSs, resulting in the following problems:

• \textbf{Low Efficiency:} In practice, it requires a long time to collect sufficient exploration data in a large-scale RS using 
existing exploration policies. This affects the speed of model iteration and means a loss of income.

• \textbf{Negative Impact on User Experience:} The exploration actions generated by existing exploration policies may include 
abnormal actions, which can negatively impact user experience and even lead to user churn—an unacceptable outcome.
\begin{figure}[hbtp!]
    \centering
    \includegraphics[width=0.72\linewidth]{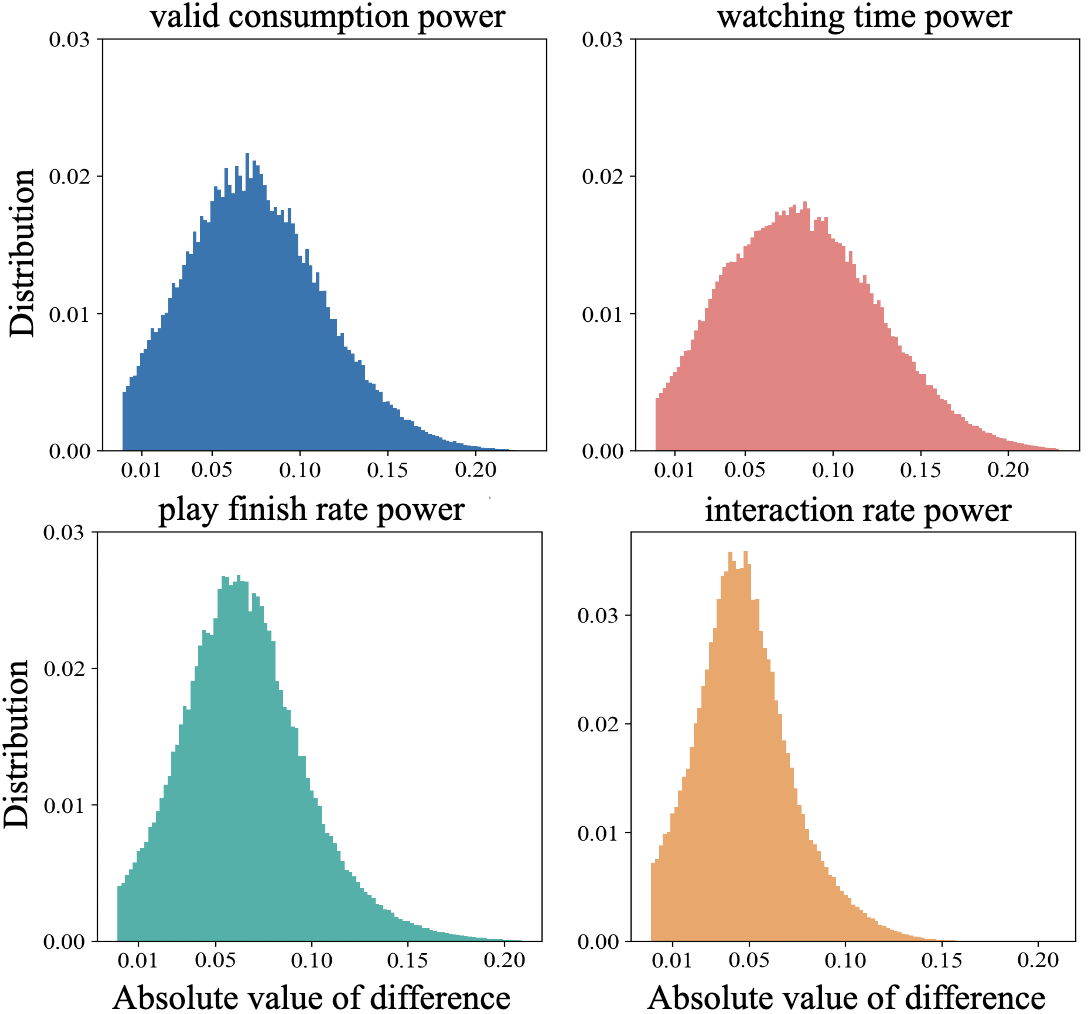}
    \caption{The distributions of absolute differences between the actions generated by the newly learned RL policy and those generated by
    the previous RL policy for the same state.}
    \label{fig:diff_dist}
\end{figure}

To solve the above issues, we first analyze the distribution of the absolute differences between the actions generated by the latest RL policy (also called actor)
and those generated by the previous RL policy for the same state. Based on this analysis, we then propose a custom exploration approach. 
It should be emphasized that, in practice, the value of each dimension of the action vector is a non-negative floating number; the below is merely a simplification for descriptive purposes.
For simplicity, we normalize the value range of each action dimension to $[-1, 1]$ and 
select the four most important action dimensions—valid consumption, watch time, play finish rate, 
and interaction behavior rate—for illustration, as shown in Figure \ref{fig:diff_dist}. 
We observe that the actions generated by the newly learned RL policy typically do not 
deviate significantly from the actions generated by the previous RL policy for the same state, 
which is also consistent with our intuition.

\begin{figure}[hbtp!]
  \centering
  \includegraphics[width=0.85\linewidth]{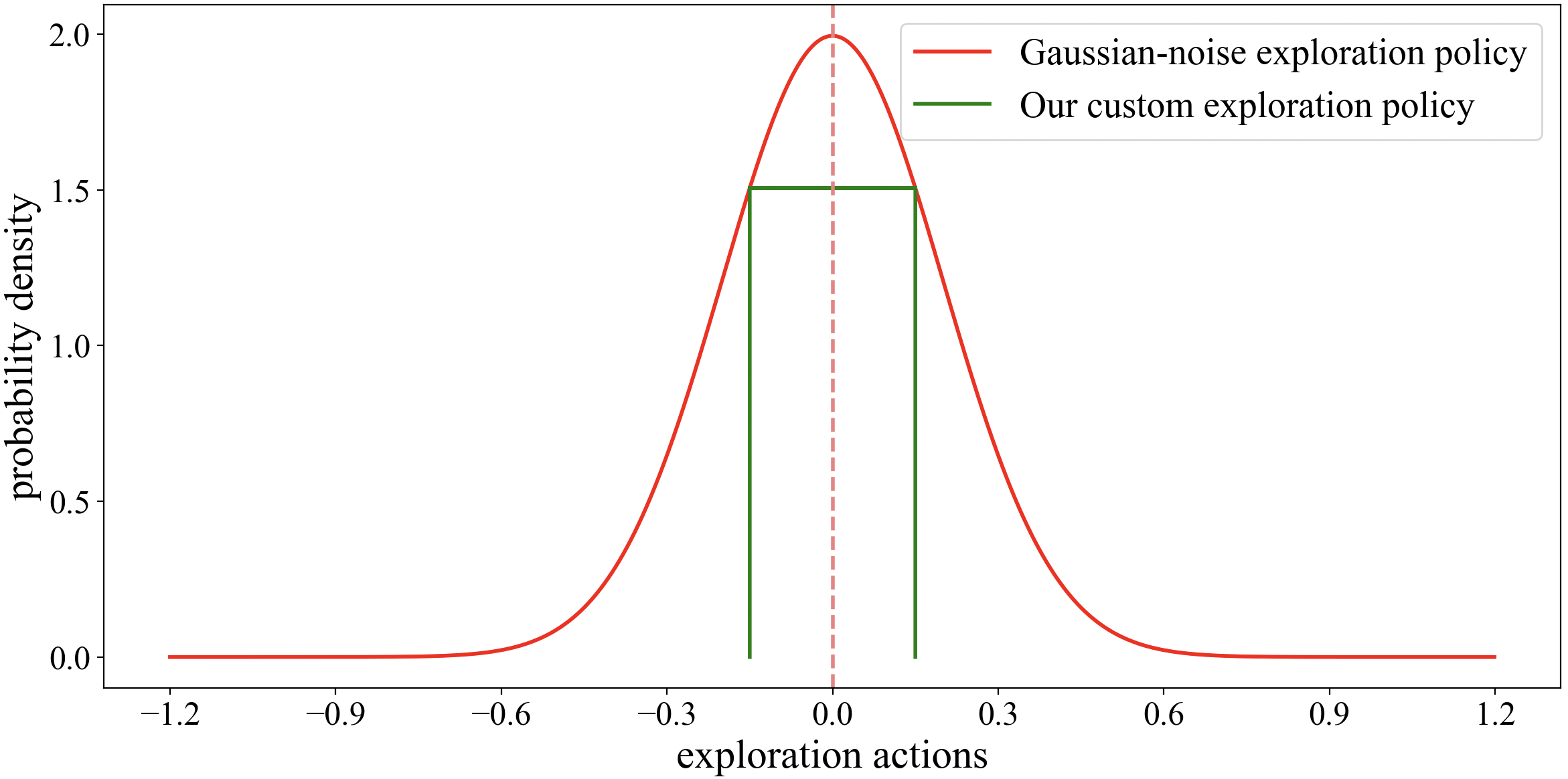}
  \caption{Comparison between the action distribution of our custom exploration policy and that of the Gaussian-noise exploration policy. 
  }
  \label{fig:explore}
\end{figure}

Inspired by this finding, we propose a simple yet highly efficient exploration policy that 
defines personalized exploration upper and lower bounds for each user, 
as shown in Eq. 4. The exploration action is generated by adding a random perturbation—sampled from a uniform distribution 
defined by $b_l$ and $b_u$—to the action outputted by the baseline policy $\mu_{bp}$ when exploring. 
Through statistical analysis, we carefully select appropriate upper and lower bounds. 
The idea of our exploration policy is to eliminate low-value exploration space and 
focus solely on exploring potential high-value state-action pairs, as illustrated in Figure \ref{fig:explore}.

\begin{equation}
    \mu _{ep}(s) = \mu_{bp}(s) + \varepsilon , \enspace \varepsilon \sim U(b_{l}, b_{u}) \,.
\end{equation}

In this way, our exploration policy demonstrates significantly higher efficiency compared to existing exploration policies. 
We take the Gaussian-noise exploration policy that is commonly used as a baseline \cite{ref15}, 
which generates exploration actions by adding Gaussian noise to the baseline policy, 
as illustrated by the red curve in Figure \ref{fig:explore}. 
Typically, we set $b_{u}$ and $b_{l}$ to $0.15$ and $-0.15$ in our recommendation scenario. 
The Gaussian-noise exploration policy previously used by us had a mean of $0.0$ and a standard deviation of $0.2$, as shown in Eq. 5. 
The action generated by our RL-MTF model is a $10$-dimensional vector. 
Therefore, under the same requirement for exploration density, the efficiency of our exploration policy is about $2^{10}$ times 
higher than that of the Gaussian-noise exploration policy. 
In addition, this approach can reduce the impact on RL model training caused by imbalanced exploration action distribution.

\begin{equation}
  \label{gauss_explore}
  \mu _{ep}(s) = \mu_{bp}(s) + \xi , \enspace \xi \sim \mathcal{N}(0.0, 0.2^2) \,.
  \vspace{0.15in}
\end{equation}

\subsection{UnifiedRL: A RL Algorithm Customized for MTF in Large-scale RSs}
\label{sec:method_details}
\begin{figure}[hbtp!]
  \centering
  \includegraphics[width=1.0\linewidth]{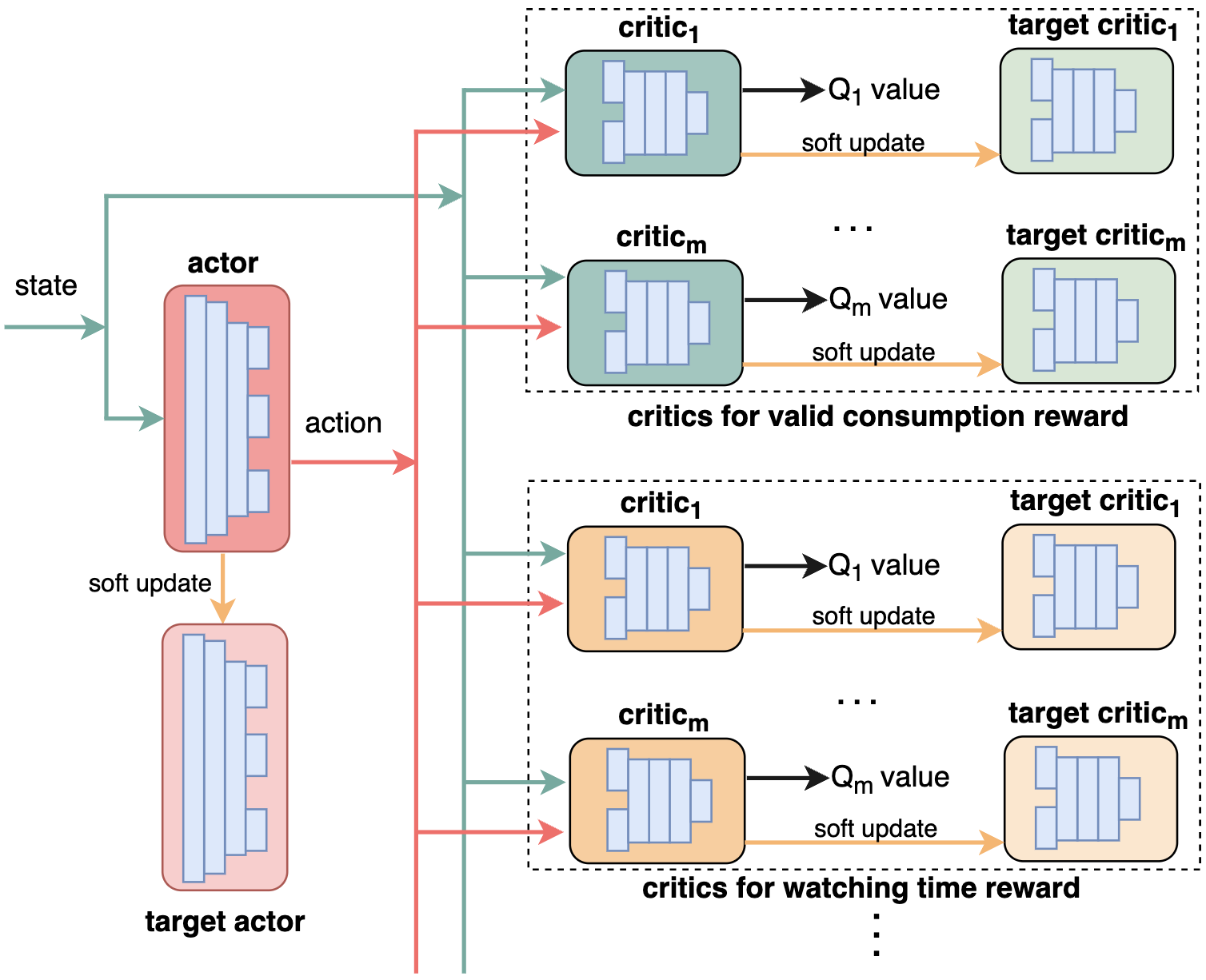}
  \caption{The framework of UnifiedRL. 
  }
  \label{fig:unified}
\end{figure}

The framework of UnifiedRL is shown in Figure \ref{fig:unified}. The key components of UnifiedRL will be introduced as below.

\subsubsection{Actor Network}
The actor network aims to output the optimal action $a_{t}$ for a given state $s_{t}$, whose structure is similar to PLE but outputs action. 
The actor and critic can also use other model structures. They are not central to UnifiedRL and are therefore not detailed description in this paper.
Following common practice, we construct two actor networks during the learning process: a current actor network $\mu(s)$ 
and a target actor network $\mu^\prime(s)$. 
The current actor network $\mu(s)$ integrates the offline policy model with its tailored efficient exploration policy within a unified framework and 
introduces an additional penalty term based on the consistency of multiple critics to mitigate the potential issue of imbalanced exploration data, 
as shown in Eq. 6 and Eq. 7. $q$ represents the number of critic sets, $q \in [1, k]$, and each set contains $m$ independent critics. 
$w_i$ is the weight of the corresponding reward and $\mu_{bp}$ denotes baseline policy during exploration.

\begin{align}
  &\theta^{k+1} \longleftarrow arg\min\limits_{\theta^{k} }E_{s_{t}\sim \mathcal{D},a\sim{\mu(s_{t}|\theta^{k})}} \;
     \Bigg[- \sum_{i=1}^q w_i \left( \bar{Q}_i^k \right) \nonumber \\
      & \qquad + \eta  d(\mu(s_{t}|\theta^{k})) + \lambda \sum_{i=1}^q w_i \left(
        \left(
          \frac{1}{m} \sum_{j=1}^m \left( Q_{ij}^k - \bar{Q}_i^k \right)^2
        \right)^{1/2}
      \right) 
    \Bigg] \,. \\ 
  & \qquad \mu_t^k = \mu(s_t \mid \theta^k); \ Q_{ij}^k = Q_{ij}(s_t,\, \mu_t^k); \ \bar{Q}_i^k = \frac{1}{m} \sum_{j=1}^m Q_{ij}^k \nonumber 
  \end{align}

  \begin{align}
    d(\mu(s_{t}|\theta^{k})) = \left\{
      \begin{array}{l} 
      0, \quad \ \ if \ \mu(s_{t}|\theta^{k}) \in (\mu_{bp}(s_{t}) + b_{l}, \mu_{bp}(s_{t}) + b_{u}) \\ \\
      \omega  \exp \left(\displaystyle\frac {\mu(s_{t}|\theta^{k}) - (\mu_{bp}(s_{t}) + b_{u})} 
      {\beta  (b_{u} - b_{l})} \right),  \qquad \qquad  (7) \\
      \qquad if \ \mu(s_{t}|\theta^{k}) \geq \mu_{bp}(s_{t}) + b_{u} \\ 
      \\ \omega  \exp \left(\displaystyle\frac{(\mu_{bp}(s_{t}) + b_{l}) - \mu(s_{t}|\theta^{k})} 
      {\beta  (b_{u} - b_{l})} \right), \\
      \qquad if \ \mu(s_{t}|\theta^{k}) \leq \mu_{bp}(s_{t}) + b_{l} \nonumber
    \end{array}
    \right.
  \end{align}

During the training of $\mu(s)$, the upper and lower bounds of the exploration data distribution for each user can be directly obtained. 
We leverage this property to relax overly strict constraints and fully exploit the capacity of $\mu(s)$. 
If the action generated by $\mu(s)$ at state $s_{t}$ falls within the user's personalized upper and lower bounds, 
the value of the second term in Eq. 6 is zero—meaning no penalty is imposed, thereby preserving the model's capacity. 
Otherwise, a penalty is applied based on the extent to which the action exceeds either the upper or lower bounds.
In this way, the performance of the current actor network is significantly improved compared to RL-MTF existing methods, 
as proved by the experiments in Section \ref{sec:experiments}. 

Furthermore, we introduce an extra penalty mechanism defined as the standard deviation of the estimated values outputted 
by multiple independent critics \cite{ref33} to alleviate potential data distribution imbalance, which corresponds to the third term in Eq. 6. 
Owing to the high efficiency of the custom exploration policy, the exploration actions collected within the user's personalized bounds 
have a significantly higher average density compared to those generated by existing exploration approaches, 
which is highly beneficial for model optimization. 
Moreover, compared with Gaussian perturbation, random perturbation within personalized upper and lower bounds reduces 
the impact of imbalanced data distribution on model training. If the action output by $\mu(s)$ lies within the user's exploration space, 
the value of the third term in Eq. 6 will be small. Otherwise, a corresponding penalty is applied.
The hyperparameters of the exploration policy need to be configured based on the specific requirements.

The target actor network $\mu^\prime(s)$ is an auxiliary network responsible for generating the next optimal action $a_{t+1}^\prime$
based on the next state $s_{t+1}$ to alleviate the overestimation problem caused by bootstrapping, 
whose parameters are periodically soft updated using the current actor network $\mu(s)$.
Other approachs to mitigate overestimation can also be used \cite{ref26, ref32}.

\subsubsection{Critic Network}
The Critic network $Q(s_{t}, a_{t})$ is responsible for estimating the cumulative reward for a given state-action pair $(s_{t}, a_{t})$ within a recommendation session. 
$Q(s, a)$ also integrates the critic network with the custom exploration policy to alleviate OOD problem. 
In UnifiedRL, a set of critics for each key reward is defined which includes $m$ independent critics, 
while unimportant rewards are merged into the key rewards, as shown in Figure \ref{fig:unified}. The number of the sets is $q$. 
All critic networks are initialized randomly and trained independently, as shown in Eq. 8 and Eq. 9. 
For example, in our recommendation scenario, the value of $q$ is set to 2. 
One set of critics is responsible for estimating the cumulative reward associated with watching time, 
while the other set focuses on the cumulative rewards of other user behaviors. Each set comprises 24 critics.
In this way, the performance of UnifiedRL can be further improved. The setting of $m$ and $q$ needs to balance performance and cost. 
If resource is sufficient, setting a larger number of them will be better.

The goal of each critic network is to minimize Temporal Difference (TD) error. 
To achieve better performance, we define a target network $Q^\prime_{ij}(s, a|\phi_{ij}^\prime)$ for each critic $Q_{ij}(s, a|\phi_{ij})$, 
whose parameters are periodically soft-updated using $Q_{ij}(s, a|\phi_{ij})$. 
If the next action generated by $\mu^\prime(s)$ at next state $s_{t+1}$ falls within the user's upper and lower bounds, 
the value of $Q_{ij}^\prime(s_{t+1}, \mu^\prime(s_{t+1})|\phi_{ij}^\prime)$ in Eq. 8 will not be punished. 
Otherwise, a penalty is applied based on the deviation that surpasses the upper or lower bounds of the user, as shown in Eq. 9. 

\begin{align}
  \phi_{ij}^{k+1} \longleftarrow	 \  &arg\min\limits_{\phi_{ij}^{k}} 
  E_{s_{t}\sim \mathcal{D}, a_{t} \sim \mathcal{D}, a_{t+1} \sim \mu^\prime(s_{t+1})} \biggl[ \bigg(Q_{ij}(s_{t}, a_{t}|\phi_{ij}^{k}) \nonumber \\ 
  &- (r_{t} + \gamma  \varphi(\mu^\prime(s_{t+1}))  Q_{ij}^\prime(s_{t+1}, \mu^\prime(s_{t+1})|\phi_{ij}^\prime))\bigg)^{2} \biggr] \,. \tag{8} \\
  & \qquad for \ i = 1, ..., q; \ j = 1, ..., m. \nonumber
\end{align}

\begin{align}
  \varphi(\mu^\prime(s_{t+1})) = \left\{
    \begin{array}{l} 
    1, \ \  if \ \mu^\prime(s_{t+1}) \in (\mu_{bp}(s_{t+1}) + b_{l}, \mu_{bp}(s_{t+1}) + b_{u}) \\ \\
    \varpi  \Biggl[ \exp \left(\zeta + \displaystyle\frac {\mu^\prime(s_{t+1}) - (\mu_{bp}(s_{t+1}) + b_{u})} 
    {b_{u} - b_{l}} \right) \Biggr]^{-1}, \ \  (9) \\
    \quad if \ \mu^\prime(s_{t+1}) \geq \mu_{bp}(s_{t+1}) + b_{u} \\ 
    \\ \varpi \Biggl[ \exp \left(\zeta + \displaystyle\frac{(\mu_{bp}(s_{t+1}) + b_{l}) - \mu^\prime(s_{t+1})} 
    {b_{u} - b_{l}} \right) \Biggr] ^{-1}, \nonumber \\
    \quad if \ \mu^\prime(s_{t+1}) \leq \mu_{bp}(s_{t+1}) + b_{l} \nonumber
  \end{array}
  \right.
\end{align}

\subsubsection{Progressive Training Mode}
One serious drawback of offline RL is that, during training, 
it relies solely on previously collected data without any further interaction with the environment. 
The absence of real-time interaction can lead to a discrepancy between the learned policy and the actual environment 
\cite{ref15, ref26, ref27, ref28, ref29, ref30, ref31}.

To alleviate this problem, we leverage UnifiedRL's efficient exploration policy to divide the previous single round of 
online exploration and offline training \cite{ref15} into multiple rounds of online exploration and offline training. 
By exploring the environment more efficiently and frequently, the target policy converges more rapidly toward the true optimal policy, 
further enhancing performance. For simplicity, we refer to this approach as the progressive training mode.

\subsection{Recommender System with RL-MTF}
\label{sec:deploy}

\begin{figure}[hbtp!]
  \centering
  \includegraphics[width=0.9\linewidth]{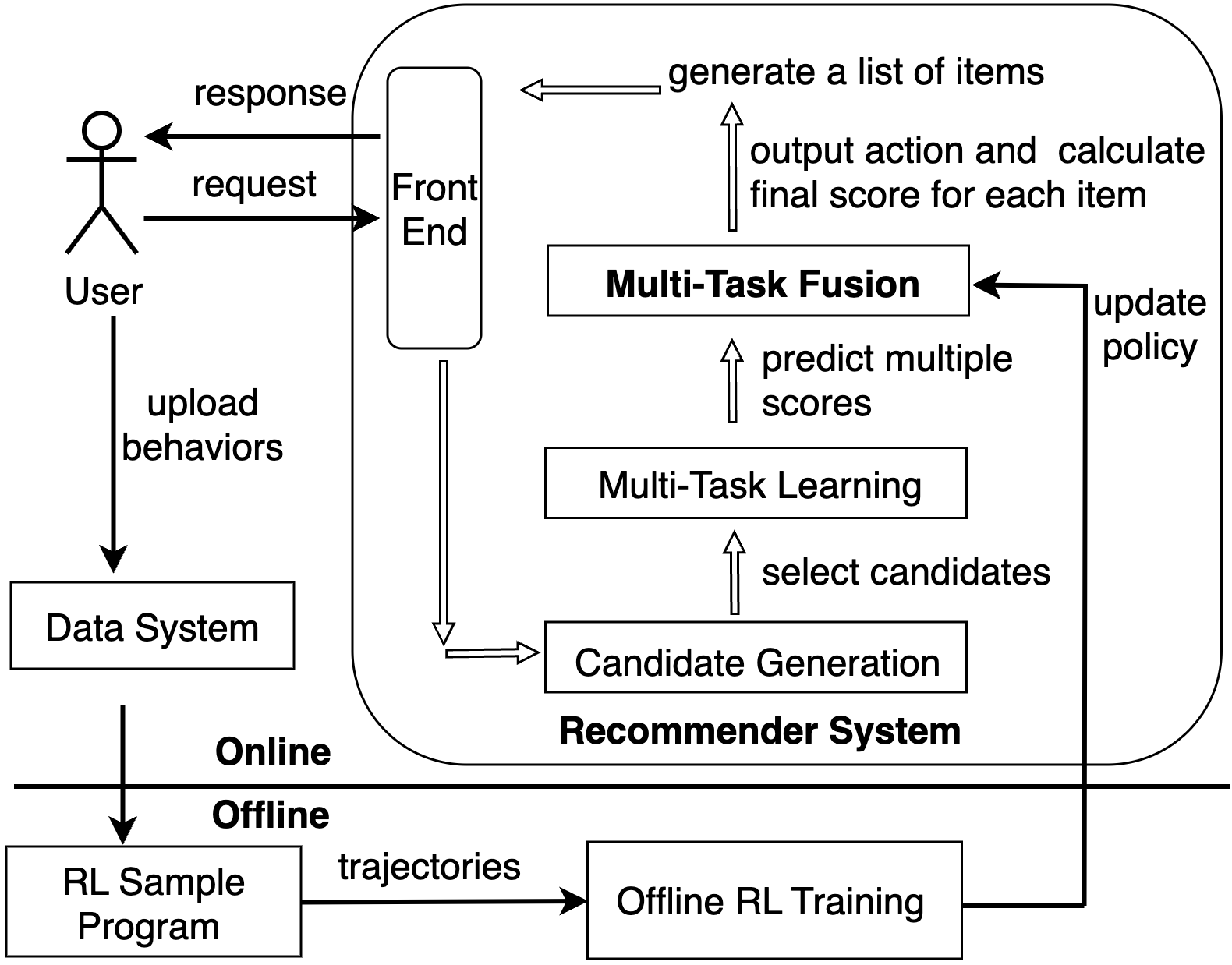}
  \caption{RL-MTF framework in recommender system.}
  \label{fig:implement}
\end{figure}

We implement UnifiedRL in a large-scale RS, as shown in Figure \ref{fig:implement}. 
The RL-MTF framework consists of two components: offline model training and online model serving.
The offline model training component is responsible for preprocessing exploration data and training the RL-MTF model.
The online model serving component primarily generates personalized optimal action when receiving a user request. 
Additionally, the online model serving component takes charge of online exploration to collect training data.


\section{Experiments}
\label{sec:experiments}

\subsection{Dataset}
\label{sec:dataset}
Since there is no publicly dataset used for MTF in RSs, we collect training data from an industrial RS, 
which serves hundreds of millions of users, according to the exploration approaches of different MTF methods. 
We use three groups of users to collect data.
Users in each group are selected randomly, with each group containing approximately 2 million users to ensure a fair comparison. 
The datasets are collected as follows:

 \textbf{Dataset 1}: It is generated using Gaussian-noise exploration policy \cite{ref15}, as shown by the red curve in Figure \ref{fig:explore}. 
In experiment, the Gaussian distribution has a mean of $0.0$ and a standard deviation of $0.2$, which were used by us previously.

 \textbf{Dataset 2}: It is produced using UnifiedRL's custom exploration policy, as shown in Eq. 4. 
 The upper and lower bounds are set to 0.15 ($b_u$) and -0.15 ($b_l$), as illustrated by the green curve in Figure \ref{fig:explore}.

 \textbf{Dataset 3}: This dataset is generated using UnifiedRL's custom exploration policy and progressive training mode, which divide 
 the above single round of online exploration and offline training into $5$ rounds of online exploration and offline training.

All the exploration policies are used to explore the environment for five days, and each dataset contains about 6.8 million sessions. 
For offline experiments, we train different RL models on the corresponding datasets and evaluate their performance. 
For online experiments, we deploy different models in a large-scale RS and conduct A/B tests.

\subsection{Compared Methods}
We compare UnifiedRL with ES and the existing RL-MTF algorithms which are commonly used in large-scale RSs. 
In addition, we design two variants of UnifiedRL to illustrate the effects of UnifiedRL model and Progressive Training Mode (PTM) 
on performance, respectively.

• \textbf{ES} \cite{ref20, ref21, ref22, ref23} uses user profile features as model input to generate personalized fusion weights. 
It is used as the benchmark for comparison.

• \textbf{BatchRL-MTF} \cite{ref15} is proposed for MTF in RSs and generates actions based on BCQ \cite{ref28}, 
which has been deployed in multiple RSs at Tencent. 
The other RL-MTF methods are implemented based on the framework proposed by BatchRL-MTF.

• \textbf{DDPG} (Deep Deterministic Policy Gradient) \cite{ref27, ref35} is a classical off-policy actor-critic algorithm that can 
learn policies in high-dimensional, continuous action spaces.

• \textbf{CQL+SAC} (Conservative Q-Learning with Soft Actor-Critic) \cite{ref30, ref39} learns a conservative, lower-bound Q function by 
regularizing the Q value of OOD action-state pairs.

• \textbf{IQL} (Implicit Q-learning) \cite{ref31} does not require evaluating actions outside of the dataset, 
but still enables the learned policy to substantially improve over the best behavior in the data through generalization.

• \textbf{UnifiedRL without PTM} is trained without the progressive training mode and 
adopts the same data exploration schedule as the above RL algorithms, in order to validate the performance of the UnifiedRL model using dataset 2.

• \textbf{UnifiedRL} is trained through frequent iterations of online exploration and offline training, with the help of its custom exploration policy, 
using dataset 3.

\subsection{Implementation Details}
\label{implement_details}
The action generated by each MTF model is a $10$-dimensional vector representing the fusion weights in Eq. 1.
All networks in the RL-MTF models are Multi-Layer Perceptrons (MLPs) and are optimized using the Adam optimizer.
The reward discount factor $\gamma$ is set to $0.9$. 
The soft update rate and the delay update step for target networks are 0.08 and 15. 
In addition, the mini-batch size and training epochs are set to $256$ and $300, 000$. 
It should be emphasized that hyperparameters need to be selected according to the specific recommendation scenario. 
The values of $\eta$, $\omega$, $\lambda$, $\beta$, $\varpi$, $\zeta$, $q$ and $m$ in UnifiedRL 
are set to 1.2, 1.0, 0.2, 0.3, 1.0, 3.0, 2 and 24, respectively. 
Through hyperparameter search, we set the hyperparameters of each model to their optimal values. 
The details of hyperparameter selection are not listed due to page limitation.

\subsection{Evaluation Setting}
\label{evaluation}
\subsubsection{Offline Policy Evaluation}
ES is the baseline service of RS and serves as the benchmark for comparison. 
DDPG, CQL+SAC, BatchRL-MTF, and IQL are trained on dataset 1. UnifiedRL without PTM is trained on dataset 2, while UnifiedRL is trained on dataset 3. 
Following \cite{ref42}, we utilize Normalized Capped Importance Sampling (NCIS) \cite{ref43} to evaluate model performance. 
The cumulative reward across all test user trajectories serves as the evaluation metric. 
Considering that NCIS requires the critic of each RL model to estimate the cumulative reward, that may introduce bias among different critics and 
can only be used for comparison between different RL models. 
Since the goal of MTF is to generate a final ranking sequence, similar evaluation metrics can be used as well. 
Inspired by this, to compare various MTF solutions, not only RL-MTF methods but also other types of MTF methods, 
we propose a new offline evaluation approach as an auxiliary evaluation metric called MTF-GAUC build on GAUC metric \cite{ref34}, 
as shown in Eq. 10.

\vspace{2mm}
\begin{equation}
  MTF\!\!-\!\!GAUC = \frac{\sum_{(u)}w_{(u)} * Weighted\_AUC_{(u)}}{\sum_{(u)}w_{(u)}} \,. \tag{10}
\end{equation}
\vspace{1mm}

To begin with, we partition all test data based on user. 
To calculate the Weighted\_AUC similar to AUC \cite{ref41} for each group, valid consumption is used as the sample label, 
the reward computed according to Eq. 2 is used as sample weight, 
and the normalization value of the final score computed according to Eq. 1 is used as prediction.
Groups that contain only positive or only negative samples are excluded from further analysis. 
Lastly, we calculate the weighted average of the Weighted\_AUC scores across all valid groups, where the weight 
$w(u)$ corresponds to the number of impressions within each group. This weighted average is taken as the MTF-GAUC value. 
Using this approach, any kind of MTF models can also be easily compared. 
In addition, it does not introduce the bias of different RL critics. 
Although it is very simple and not rigorous, it works well in practice.

\subsubsection{Online A/B Testing}
We evaluate each model using user valid consumption and user duration time, 
which are the two most important online metrics in our recommendation scenario.

\textbf{User Valid Consumption (UVC):} The average total valid consumptions per user in a day. 
A valid consumption is defined as a user watching a video for more than $10$ seconds.

\textbf{User Duration Time (UDT):} The average total watching time per user within a day.

\subsection{Offline Evaluation}
\subsubsection{Effectiveness of UnifiedRL}
To compare the performance of the above MTF algorithms, we train their models separately and evaluate them using both NCIS and MTF-GAUC, 
as shown in Table \ref{table:offline_auc1} and Table \ref{table:offline_auc2}. ES is taken as benchmark and is not an RL model, 
therefore NCIS is not used to evaluate ES. The results of these two evaluation approaches are consistent.
\begin{table}[hbtp!]
    \caption{The cumulative reward of the compared RL-MTF methods.}
    \label{table:offline_auc1}
    \footnotesize
    \renewcommand\arraystretch{1.1}{
    \begin{minipage}{\columnwidth}
    \begin{center}
    \resizebox{\linewidth}{!}{
    \small{
    \begin{tabular}{ c|c }
    \toprule
    \textbf{ \qquad Compared Methods \qquad  } & \textbf{\quad \ \ Cumulative Reward \ \ \quad } \\ \hline
    DDPG & 51.62 \\ 
    CQL+SAC & 51.85 \\ 
    BatchRL-MTF & 52.09 \\ 
    IQL & 52.39 \\ 
    \textbf{ UnifiedRL without PTM  } & \textbf{ 53.82} \\ 
    \textbf{UnifiedRL} & \textbf{\ 53.96} \\
    \bottomrule
    \end{tabular}
    }}
    \end{center}
    \end{minipage}
    }
\end{table}

In offline evaluation, the MTF-GAUC of DDPG is higher than that of ES model, 
due to its more powerful performance and consideration of long-term rewards, as shown in Table \ref{table:offline_auc2}
The cumulative reward of UnifiedRL without PTM is significantly higher than that of the existing offline RL models, 
as shown in Table \ref{table:offline_auc1}. 
As previously mentioned, to avoid OOD, existing offline RL algorithms impose overly strict constraints, 
which significantly impairs their performance. This is because these algorithms are limited to training on a fixed dataset and 
are unaware of the exploration policy that generated the data. 
Therefore, they can only avoid OOD problem through excessively strict constraints.
In RSs, we design a custom online exploration policy for our offline RL algorithm, 
and during offline model training, the upper and lower bounds of the exploration data distribution for each user can be directly obtained. 
We leverage this characteristic to simplify the overly strict constraints and 
integrate the offline model algorithm with our efficient exploration policy. 
In this way, UnifiedRL without PTM significantly improves its model performance. 
Furthermore, UnifiedRL outperforms UnifiedRL without PTM. 
Through multiple iterations of environment exploration with our efficient exploration policy, the learned policy is progressively enhanced.
\begin{table}[hbtp!]
    \caption{The MTF-GAUC of the compared methods on the same test dataset.}
    \label{table:offline_auc2}
    \footnotesize
    \renewcommand\arraystretch{1.1}{
    \begin{minipage}{\columnwidth}
    \begin{center}
    \resizebox{\linewidth}{!}{
    \small{
    \begin{tabular}{ c|c }
    \toprule
    \textbf{\quad  Compared Methods \quad } & \textbf{\qquad \   MTF-GAUC \   \qquad } \\ \hline
    ES & 0.7836 \\ 
    DDPG & 0.7881 \\ 
    CQL+SAC & 0.7892 \\ 
    BatchRL-MTF & 0.7894 \\ 
    IQL & 0.7906 \\ 
    \textbf{ \qquad UnifiedRL without PTM \qquad  } & \textbf{ 0.7941 } \\ 
    \textbf{UnifiedRL} & \textbf{0.7953} \\ 
    \bottomrule
    \end{tabular}
    }}
    \end{center}
    \end{minipage}
    }
  \end{table}

\subsection{Online Evaluation}
The results of online experiments in large-scale RSs are the ultimate criterion for evaluating the effectiveness of a method. 
So we deploy all the compared models in an industrial RS for one week to conduct A/B tests. 
ES is used as the baseline to demonstrate the improvements of the other models. 
The results of the online experiments are shown in Table \ref{table:online_res}, and all improvements are statistically significant with 
$p$-values less than $0.05$.

DDPG increases user valid consumption by $+1.39\%$ and user duration time by $+0.81\%$ compared to ES, 
benefiting from its much stronger model performance and consideration of long-term rewards. 
CQL+SAC outperforms DDPG, and BatchRL-MTF slightly outperforms CQL+SAC. 
IQL surpasses BatchRL-MTF, achieving a $+2.09\%$ increase in user valid consumption and a $+1.15\%$ 
increase in user duration time compared to ES. 
UnifiedRL significantly outperforms all other algorithms, with a $+4.64\%$ increase in user valid consumption and 
a $+1.74\%$ increase in user duration time compared to the benchmark. 
Moreover, UnifiedRL has also been applied to several other large-scale RSs in our company, achieving significant improvements.
\begin{table}[hbtp!]
    \caption{The online results of the compared methods in our large-scale RS}
    \label{table:online_res}
    \footnotesize
    \renewcommand\arraystretch{1.1}{
    \begin{minipage}{\columnwidth}
    \begin{center}
    \resizebox{\linewidth}{!}{
    \small{
    \begin{tabular}{ c|c|c }
    \toprule
    \textbf{\quad Compared Methods \quad} & \textbf{\qquad UVC \qquad} & \textbf{\qquad UDT \qquad} \\ \hline
    ES & * & * \\ 
    DDPG & +1.39\% & +0.81\%  \\ 
    CQL+SAC & +1.62\% & +0.95\% \\ 
    BatchRL-MTF & +1.79\% & +0.98\% \\ 
    IQL & +2.09\% & +1.15\% \\ 
    \textbf{\qquad UnifiedRL \qquad } & \textbf{ +4.64\%  } & \textbf{ +1.74\% } \\
    \bottomrule
    \end{tabular}
    }}
    \end{center}
    \end{minipage}
    }
\end{table}

\section{Conclusion}
\label{conclusion}
In this work, we first point out the problems of existing RL-MTF studies and then propose UnifiedRL tailored for large-scale RSs.
UnifiedRL integrates offline RL algorithm with its custom online exploration policy through utilizing the characteristics of RSs. 
which significantly improves performance. 
Furthermore, we employ progressive training mode to learn the optimal policy with the help of the custom exploration policy, 
further improving model performance. 
Offline experiments show that UnifiedRL significantly exceeds other existing MTF methods. 
And the online results conducted in an industrial RS demonstrate that UnifiedRL achieves remarkable improvements over other models. 
Up to now, UnifiedRL has been applied in multiple large-scale RSs and has also been adopted in search engines and advertising systems.

\section*{GenAI Usage Disclosure}
I know that the ACM's Authorship Policy requires full disclosure of all use of generative AI tools in all stages of the research 
(including the code and data) and the writing. No GenAI tools were used in any stage of the research, nor in the writing.


\end{document}